\documentstyle{l-aa}     
\def\ref{\par \noindent \hang}

\def\etal{et al.\thinspace}
\def\eg{{\it e.g.\ }}

\def\ie{{\it i.e.\ }}
\def\approxlt{\mathrel{\hbox{ \lower .5ex \hbox {$\sim$}
	\llap{\raise .15 ex \hbox{$<$}} }}}
\def\approxgt{\mathrel{\hbox{ \lower .5ex \hbox {$\sim$}
	\llap{\raise .15 ex \hbox{$>$}} }}}

\def\multleft#1{\hbox to size{\vbox {\halign {\lft{##}\cr #1}}\hfill}\par}
\def\multright#1{\hbox to size{\vbox {\halign {\rt{##}\cr #1}}\hfill}\par}

\def\today{\ifcase\month\or January\or February\or March\or April\or May\or
      June\or July\or August\or September\or October\or November\or December\fi
      \space\number\day, \number\year}
\def\<{\thinspace}
\def\s{\hbox{\phantom{5}}}      

\def\boxit#1{\vbox{\hrule\hbox{\vrule\kern3pt\vbox{\kern3pt
	  #1 \kern3pt}\kern3pt\vrule}\hrule}}

\def\km{{\rm\thinspace km}}

\def\Mpc{{\rm\thinspace Mpc}}

\def\s{{\rm\thinspace s}}

\def\kmps{\hbox{$\km\s^{-1}\,$}}

\def\kmpspMpc{\hbox{$\kmps\Mpc^{-1}$}}

\def\H2{\hbox{H$_{2}$~}}

\begin{document}
 
\input psfig.sty
\thesaurus{03(11.01.2; 11.02.1; 11.14.1; 11.16.1; 13.09.1)}

\title{The host galaxies of BL Lac objects in the near--infrared\thanks{Based 
on observations collected at the European Southern Observatory, La Silla, 
Chile.}}

\author{Jari K. Kotilainen \inst{1,2}, Renato Falomo\inst{3} and Riccardo 
Scarpa\inst{4}}
\offprints{J.K. Kotilainen (Tuorla address)}

\institute{International School for Advanced Studies (SISSA), via Beirut 
2--4, I--34014 Trieste, Italy
\and
Tuorla Observatory, University of Turku, V\"{a}is\"{a}l\"{a}ntie 20, 
FIN--21500 Piikki\"{o}, Finland; e--mail: jarkot@deneb.astro.utu.fi
\and
Osservatorio Astronomico di Padova, vicolo dell'Osservatorio 5, I--35122 
Padova, Italy; e--mail: falomo@astrpd.pd.astro.it
\and
Space Telescope Science Institute, 3700 San Martin Drive, Baltimore, MD 
21218, U.S.A; e--mail: scarpa@stsci.edu}

\date{Accepted May 18, 1998; received April 10, 1998; in original form 
February 21, 1998}

\maketitle

\markboth{J.K. Kotilainen, R. Falomo \& R. Scarpa: BL Lac hosts in the NIR}{}

\begin{abstract}

We present the results of near--infrared H band (1.65 $\mu$m) imaging of 11 
BL Lac objects with redshifts ranging from z = 0.05 to 0.9. We are able to 
clearly detect the host galaxy in seven low redshift (z$\leq$0.24) BL Lacs, 
while the four unresolved BL Lacs have either high or unknown redshift. The 
galaxies hosting the low redshift BL Lacs are large (average bulge scale 
length R(e) = 8.8$\pm$9.9 kpc) and luminous (average M(H) = --25.8$\pm$0.5), 
\ie slightly brighter than the typical galaxy luminosity L* (M*(H) = 
--25.0$\pm$0.2), and of similar luminosity to or slightly fainter than 
brightest cluster galaxies (M(H) = --26.3$\pm$0.3). The average 
optical/near--infrared colour and colour gradient of the BL Lac hosts (R--H = 
2.2$\pm$0.5; $\Delta$(R--H)/$\Delta$(log r) = --0.09$\pm$0.04) are consistent 
with the hosts being normal ellipticals, indicating that the nuclear activity 
has only a marginal effect on the star formation history and other properties 
of the hosts. The BL Lac hosts appear slightly less luminous than those of 
higher redshift flat spectrum radio quasars. The nucleus--to--galaxy 
luminosity ratio of the BL Lacs is similar to that of low redshift radio 
galaxies and consistent with what found in previous optical studies of BL 
Lacs. However, it is smaller that that found for flat spectrum radio quasars, 
suggesting there is a difference in the intrinsic brightness of the nuclear 
source or in the Doppler beaming factor between the two types of blazars.


\end{abstract}

\section{Introduction}

BL Lac objects are active galactic nuclei (AGN) characterized by strong and 
rapidly variable continuum emission and polarization across the 
electromagnetic spectrum, strong compact flat spectrum radio emission and 
superluminal motion (see \eg Kollgaard \etal 1992 for general references). 
They share many properties with flat spectrum radio quasars (FSRQ) and are 
often grouped together as blazars. The clearest difference between them is 
that the latter have strong broad emission lines, while these are very weak 
or absent in BL Lacs. 

Blazar properties are usually explained by the beaming model (Blandford \& 
Rees 1978), where the observed emission is dominated by a synchrotron 
emitting relativistically boosted jet oriented close to our line--of--sight. 
This model is supported by the fact that almost all blazars are strong and 
rapidly variable $\gamma$--ray sources (\eg von Montigny \etal 1995). The 
beaming model implies the existence of a more numerous parent population of 
intrinsically identical objects but with their jet oriented at larger angles 
to our line--of--sight. In the current unified models for radio--loud AGN 
(\eg Urry \& Padovani 1995), BL Lac objects are unified with low luminosity 
core--dominated (F--R I) radio galaxies (RG) seen nearly along the jet axis, 
while the high luminosity lobe--dominated (F--R II) RGs represent the parents 
of FSRQs (Padovani \& Urry 1990; Urry, Padovani \& Stickel 1991). However, 
for potential problems in this simple unification, see Urry \& Padovani 
(1995). For a direct test of the unification model, we need to compare 
orientation--independent properties of BL Lac objects with those of the 
parents, \eg extended radio emission, host galaxies and environments. 

Considerable amount of optical imaging exists for relatively nearby 
(z$\leq$0.5) BL Lac hosts (\eg Abraham, McHardy \& Crawford 1991; Stickel, 
Fried \& K\"{u}hr 1993; Falomo 1996; Wurtz \etal 1996; Falomo \etal 1997b; 
Jannuzi, Yanny \& Impey 1997). The host galaxies of nearby BL Lacs have 
turned out to be predominantly giant ellipticals with similar magnitude to 
F--R I RGs, although there appear to be some cases of disk dominated host 
galaxies (\eg McHardy \etal 1991; Abraham \etal 1991; Stocke, Wurtz \& 
Perlman 1995, but see contradicting views by \eg Romanishin 1992; Stickel 
\etal 1993; Falomo \etal 1997a). The extended radio power and optical 
environments of BL Lacs are also consistent with those of F--R I RGs, but 
suggest a contribution to the parent population from F--R II RGs (Kollgaard 
\etal 1992; Pesce, Falomo \& Treves 1995; Wurtz \etal 1997). 

Very little near--infrared (NIR) imaging exists on BL Lac objects. However, 
NIR wavelengths may offer some advantages. Optical emission from BL Lacs is 
often dominated by the nuclear source, while the luminosity of the massive 
old stellar population peaks in the NIR. This leads to a better contrast of 
the nebulosity with respect to the nuclear source at these wavelengths. One 
also needs to apply much lower K--correction in the NIR than in the optical. 
In this paper we present NIR H--band (1.65 $\mu$m) images of 11 BL Lac 
objects and compare the NIR host properties with those of RGs and FSRQs. The 
BL Lacs were observed during our project to study the host galaxies of a 
complete sample of FSRQs (Kotilainen, Falomo \& Scarpa 1998; hereafter KFS98) 
and thus they do not satisfy any criteria of completeness. However, all the 
low redshift BL Lacs in this sample have previously been imaged in the 
optical by us. The same procedure of analysis was performed on the NIR and 
optical datasets, thus allowing us to investigate the R--H colour of the host 
galaxies in a homogeneous manner. Properties of the observed objects are 
given in Table 1. In section 2, we briefly describe the observations, data 
reduction and the method of the analysis and refer the reader to a more 
thorough discussion given in KFS98. Our results are presented in section 3 
and conclusions in section 4. Throughout this paper, H$_{0}$ = 50 \kmpspMpc 
and q$_{0}$ = 0 are used. 

\begin{table*}
\begin{center}
\begin{tabular}{llrllrrr}
\multicolumn{8}{c}{{\bf Table 1.} Journal of observations.}\\
\hline\\
\multicolumn{1}{c}{Name} & Other name & z & V & M(B) & Date & Exp. time & 
FWHM \\
\medskip
     & 	          &   &   &      &      & (min)    & (arcsec) \\
\hline\\
PKS 0048--097 & OB-080   & $\geq$0.5(?) & 16.3 & -- & 21/8/95 & 40 & 1.0 \\
PKS 0118--272 & OC-230.4 & $\geq$0.557 & 15.9 & $\leq$-26.8 & 18/8/95 & 37 & 
1.0 \\
PKS 0521--365 &          & 0.055   & 14.6 & -22.3 & 13/1/96 & 21 & 1.2 \\
PKS 0537--441 &          & 0.896 & 15.0 & (-27.0) & 13/1/96 & 36 & 1.1 \\
PKS 0548--322 &          & 0.069   & 15.5 & -22.0 & 13/1/96 & 28 & 1.0 \\
PKS 1514--241 & AP Lib   & 0.049   & 14.9 & -21.7 & 19/8/95 & 15 & 0.9 \\
PKS 1538$+$149  & 4C 14.60 & 0.605   & 17.8 & -25.2 & 21/8/95 & 40 & 1.7 \\
PKS 2005--489 &          & 0.071   & 14.4 & -24.8 & 19/8/95 & 30 & 1.0 \\
MS 2143.4$+$070 &          & 0.237   & 18.0 & -22.8 & 18/8/95 & 40  & 0.9\\
PKS 2155--305 &          & 0.116 & 13.5 & -25.9 & 19/8/95 & 22 & 1.0 \\
PKS 2254$+$074  & OY 091   & 0.190   & 16.4 & -23.3 & 18/8/95 & 10 & 1.2 \\
\hline\\
\end{tabular}
\end{center}
\end{table*}

\section{Observations, data reduction and modeling of the luminosity profiles}

The observations were carried out at the 2.2m telescope of European Southern 
Observatory (ESO), La Silla, Chile, in August 1995 and January 1996, using 
the 256x256 px IRAC2 NIR camera (Moorwood \etal 1992) with pixel scale 
0.27$''$ px$^{-1}$, giving a field of view of 69 arcsec$^2$. Log of the 
observations is given in Table 1. The observations, data reduction and 
modeling of the luminosity profiles were performed following the procedure 
described in KFS98. Briefly, for nearby targets object frames were 
interspersed with sky frames. Distant targets, on the other hand, were always 
kept in the field by shifting them across the array. Individual exposures 
were coadded to achieve the final integration time (see Table 1).

Data reduction consisted of correction for bad pixels by interpolating across 
neighboring pixels, sky subtraction using a median averaged sky frame (for 
nearby sources) or a median averaged frame of all observations (for distant 
sources), flat-fielding, and combination of images of the same target. 
Standard stars from the list of Landolt (1992) were used for photometric 
calibration, for which we estimate an accuracy of $\sim$0.1 mag. 
K--correction was applied to the host galaxy magnitudes following the method 
of Neugebauer \etal (1985). The size of this correction is insignificant at 
low redshift in the H filter (m(H) = 0.01 at z = 0.2).
No K--correction was applied to the nuclear component, assumed to have a 
power-law spectrum ($f_\nu \propto \nu^{-\alpha}$) with $\alpha$ $\sim$--1.

Azimuthally averaged radial luminosity profile was extracted for each object 
and stars in the frames down to surface brightness of $\mu$(H) = 22--23 mag 
arcsec$^{-2}$. Only for few objects, bright stars were present in the field 
for straightforward PSF determination. For most sources, the core of the PSF 
was derived from faint field stars, while the wing was extrapolated using a 
suitable Moffat (1969) function obtained from fitting bright stars in other 
frames of similar seeing during the same night. We stress that this 
extrapolation does not affect the fit for the low redshift, often 
host--dominated BL Lacs in our sample. The luminosity profiles were fitted 
into point source and galaxy components by an iterative least-squares fit to 
the observed profile. For low redshift objects, we attempted both elliptical 
(de Vaucouleurs law) and exponential disc models to represent the galaxy. 
However, consistently with results from optical images, in no case did the 
disc model give a better fit than the elliptical one for the observed 
sources. For higher redshift objects, data quality does not allow one to 
discriminate between the morphologies and for them the elliptical model was 
assumed. We estimate the uncertainty of the derived host galaxy magnitudes to 
be $\sim$$\pm$0.3 mag. 

\section{Results and discussion}

In Fig. 1 we show the H band contour plots of all the BL Lacs, after 
smoothing the images with a Gaussian filter of $\sigma$ = 1 px. We are able 
to clearly detect the host galaxy in all the BL Lacs at low redshift 
(z$\leq$0.24). In  the unresolved cases, the redshift is either high (PKS 
1538+14; z = 0.605, PKS 0537--441; z = 0.896) or unknown but probably high 
(z$\geq$0.5; PKS 0048--097 and z$\geq$0.557; PKS 0118--272). Therefore, these 
sources are still consistent with being hosted by giant ellipticals. On the 
other hand, if their redshifts are lower, they may reside in sub--luminous 
hosts, which could account for the failure to detect absorption lines in the 
spectra of the latter two BL Lacs. In Fig. 2, we show the radial luminosity 
profiles of each BL Lac object, together with the best--fit models overlaid. 
In the Appendix, we compare our NIR photometry with previous studies, and 
discuss in detail individual BL Lacs, including comparison with previous 
optical determinations of the host galaxies. The results derived from the 
profile fitting are summarized in Table 2, where column (1) and (2) give the 
name and redshift of the object; (3) the bulge scalelength in arcsec and kpc; 
(4) and (5) the apparent nuclear and host galaxy magnitude; (6) the 
nucleus/galaxy luminosity ratio; (7) and (8) the absolute nuclear and host 
galaxy magnitude; and (9) whether the image of object is resolved (R) or 
unresolved (U). 

\begin{figure*}
\psfig{file=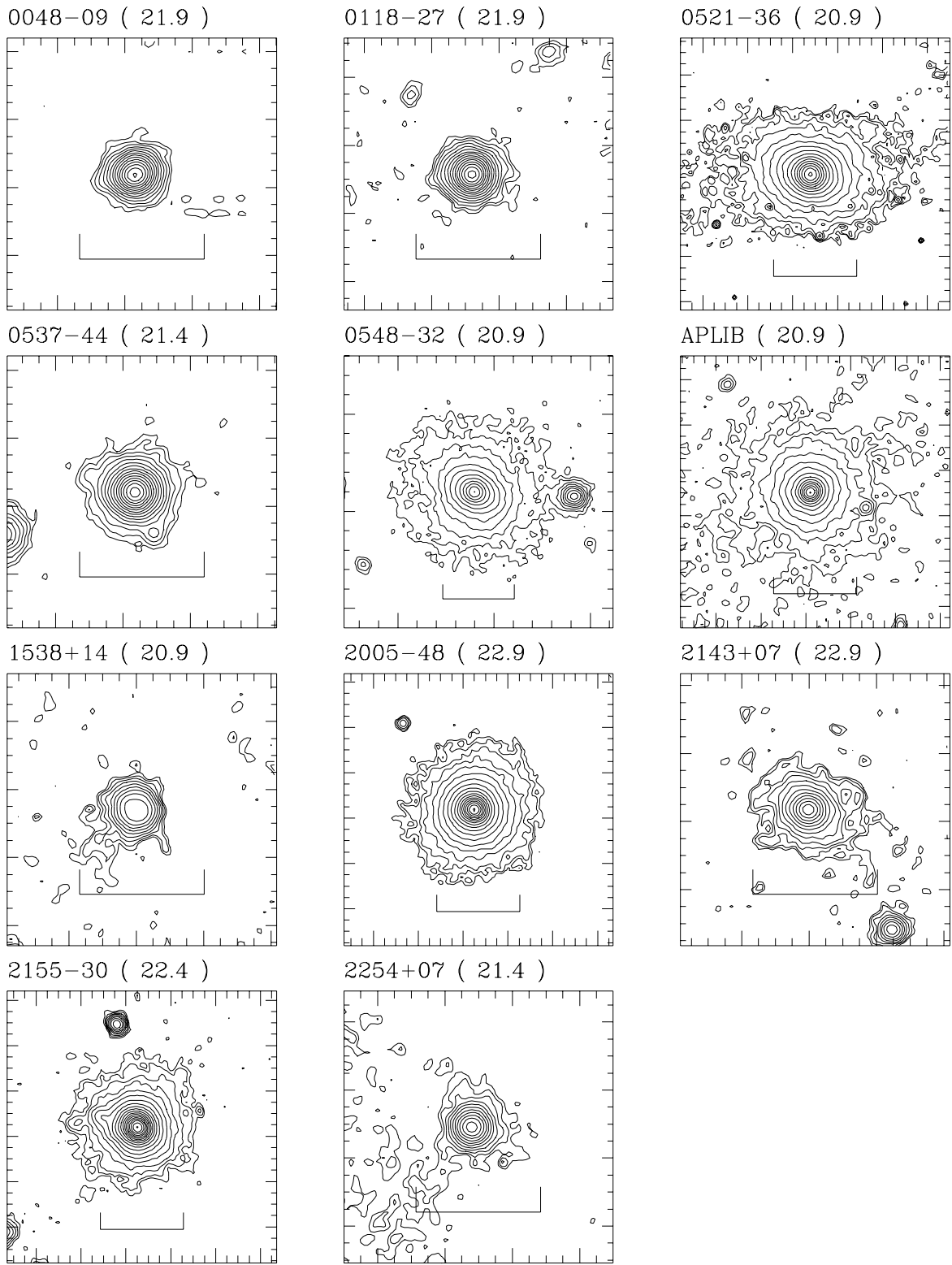,width=18cm,height=24.3cm}
\caption{\label{fig:fig1} 
Gaussian smoothed contour plots of the sample objects in the H band.  The 
scale of each image is indicated by a 10 arcsec bar.  The surface brightness 
of the lowest contour is indicated in parenthesis. The contours are separated 
by 0.5 mag intervals. North is up and east to the left.}
\end{figure*}

\begin{figure*}
\psfig{file=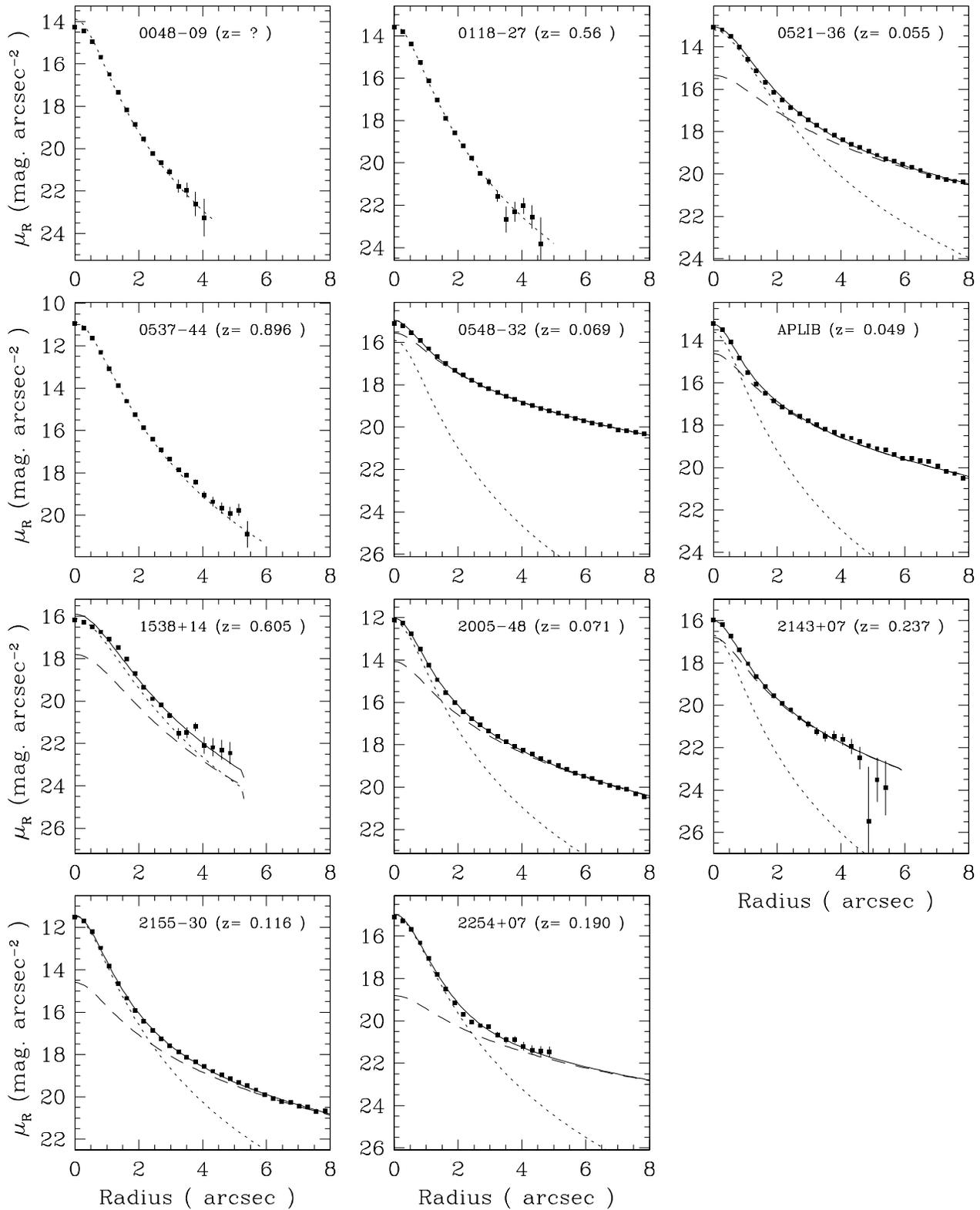,width=18cm,height=23.8cm}
\caption{\label{fig:fig2} 
Results of the profile fits for each galaxy. The solid points represent the 
observed profile, dotted line the PSF, dashed line the de Vaucouleurs model 
convolved with the proper PSF, and the solid line the fitted model profile.}
\end{figure*}

\begin{table*}
\begin{center}
\begin{tabular}{lllllllll}
\multicolumn{9}{c}{{\bf Table 2.} Properties of the host galaxies.}\\
\hline\\
\multicolumn{1}{c}{Name} & z & r(e)/R(e) & m(nuc) & m(g) & L(nuc)/L(gal) & 
M(nuc) & M(gal) & Note\\
         &   & arcsec/kpc &       &      &       &        &       & \\
(1) & (2) & (3) & (4) & (5) & (6) & (7) & (8) & (9) \\
\hline\\
PKS 0048--097  & $\geq$0.5(?) & & 13.7 & & $\geq$100 & $\leq$-29.2 & & U\\
PKS 0118--272  & $\geq$0.557 & & 13.1 & & $\geq$100 & $\leq$-30.1 & & U\\
PKS 0521--365 & 0.055 & 2.55/3.8 & 12.0 & 12.4 & 1.70 & -25.8 & -25.2 & R\\
PKS 0537--441  & (0.896) & & 13.0 & & $\geq$100 & (-32.0) & & U\\
PKS 0548--322  & 0.069 & 4.55/8.3 & 15.4 & 12.4 & 0.087 & -23.0 & -25.7 & R \\
PKS 1514--241 & 0.048 & 2.50/3.3 & 13.3 & 12.2 & 0.43 & -24.1 & -25.1 & R\\
PKS 1538$+$149 & 0.605 & & 14.7 & & 3.0 & -28.7 & & U \\
PKS 2005--489 & 0.071 & 1.65/3.1 & 11.7 & 11.9 & 1.38 & -26.6 & -26.3 & R\\
MS 2143.4$+$070 & 0.237 & 1.10/5.5 & 16.4 & 15.1 & 0.35 & -24.7 & -25.9 & R\\
PKS 2155--305 & 0.116 & 1.75/5.0 & 11.0 & 12.4 & 4.00 & -28.5 & -26.9 & R\\
PKS 2254$+$074 & 0.190 & 7.60:/32.5: & 14.0: & 14.9: & 2.27: & -26.5: & 
-25.6: & R \\
\hline\\
\multicolumn{9}{l}{$:$ uncertain values (see Appendix).}\\
\end{tabular}\\
\end{center}
\end{table*}

\subsection{The near--infrared properties of the host galaxies}

The host galaxies of all the low redshift resolved BL Lacs in this sample 
have previously been studied by us in the optical (Falomo 1996; Falomo \& 
Kotilainen, in preparation). The observations presented here are, however, 
the first study of them in the NIR, and allow us to address for the first 
time the issue of the optical--NIR colour of the BL Lac host galaxies. The 
integrated colours of the host galaxies are given in Table 3, where column 
(1) gives the name of the BL Lac; (2) and (3) the absolute magnitude of the 
host galaxy in the H band (this work) and R band (literature value), 
respectively; (4) the reference for column (3); (5) the R--H colour of the 
host galaxy computed from columns (2) and (3); (6) the same as column (5), 
but computed from our colour profiles (see below); and (7) the R--H colour 
gradient of the host. For comparison, in Table 4 we give optical and 
optical--NIR colours of elliptical galaxies from literature search. The 
average and median R--H host colours of our BL Lac sample, R--H = 2.2$\pm$0.5 
and R--H = 2.3 are consistent with those found for local giant ellipticals 
(R--H = 2.5; see Table 4), considering the large variations both within the 
sample and in the previous optical determinations of the host parameters. The 
average R--H colour also agrees well with recent evolutionary population 
synthesis models, which predict for low redshift, solar metallicity 
ellipticals R--H$\sim$2.2 (Poggianti 1997; age 15 Gyr) and R--H$\sim$2.4 
(Fioc \& Rocca-Volmerange 1997; age 13 Gyr). 

\begin{table*}
\begin{center}
\begin{tabular}{lllllll}
\multicolumn{7}{c}{{\bf Table 3.} Optical--NIR colours of the host galaxies.}\\
\hline\\
\multicolumn{1}{c}{Name} & M(H)  & M(R)  & Ref. & R-H & R-H & 
$\Delta$(R-H)/$\Delta$(log r)\\
(1) & (2) & (3) & (4) & (5) & (6) & (7) \\
\hline\\
PKS 0521--365  & -25.4 & -23.2 & F94  & 2.2 & 2.3 & -0.08\\
           &       & -23.0 & W96  & 2.4 & & \\
PKS 0548--322  & -25.7 & -24.2 & F95  & 1.5 & 2.1 & -0.06\\
           &       & -23.2 & W96  & 2.5 & & \\
PKS 1514--241  & -25.1 & -22.8 & B87  & 2.3 & 2.1 & -0.12\\
           &       & -22.8 & A91  & 2.3 & & \\
           &       & -23.5 & S93  & 1.6 & & \\
           &       & -22.9 & this work & 2.2 & & \\
PKS 2005--489  & -26.3 & -24.2 & S93  & 2.1 & 2.7 & -0.15\\
           &       & -23.7 & F96  & 2.6 & & \\
MS 2143.4$+$070 & -25.9 & -23.4 & W96  & 2.5 & 2.5 & -0.04\\
PKS 2155--305  & -26.8 & -24.4 & F96  & 2.4$^a)$ & & \\
PKS 2254$+$074   & -25.6: & -24.1 & S93  & 1.5: & 1.8: & 0.10: \\
           &       & -23.9 & F96  & 1.7: & & \\
           &       & -23.9 & W96  & 1.7: & & \\
\hline\\
\multicolumn{7}{l}{$^a)$ Original observations obtained in the Gunn i filter 
by F91.}\\
\multicolumn{7}{l}{$:$ uncertain values (see Appendix).}\\
\multicolumn{7}{l}{A91 = Abraham \etal (1991), B87 = Baxter, Disney \& 
Phillipps (1997),}\\
\multicolumn{7}{l}{F94 = Falomo (1994), F95 = Falomo, Pesce \& Treves (1995), 
F96 =}\\
\multicolumn{7}{l}{Falomo (1996), S93 = Stickel \etal (1993), W96 = Wurtz 
\etal (1996).}\\
\end{tabular}\\
\end{center}
\end{table*}

\begin{table*}
\begin{center}
\begin{tabular}{lllllll}
\multicolumn{7}{c}{{\bf Table 4.} Optical and optical--NIR colours of 
elliptical galaxies from literature.}\\
\hline\\
\multicolumn{1}{c}{Ref.} & V--R & V--I & V--K & J--H & H--K & J--K \\
\hline\\
Gregg (1989) & 0.58$\pm$0.11 & 1.23$\pm$0.22 & & & & \\
Carollo \etal (1997) & & 1.30$\pm$0.05 & & & & \\ 
Schombert \etal (1993) & & & 3.29$\pm$0.09 & & & 0.87$\pm$0.04 \\ 
Bressan, Chiosi \& Tantalo (1996) & & & 3.28$\pm$0.11 & & & \\ 
Recillas-Cruz \etal (1990) & & & & 0.72$\pm$0.08 & 0.22$\pm$0.06 & \\ 
Silva \& Elston (1994) & & & & & & 0.90$\pm$0.02 \\ 
\hline\\
\end{tabular}\\
\end{center}
\end{table*}

We have next computed radial R--H colour profiles for six of the BL Lacs 
(Fig. 3). From these profiles we have computed the R--H colour gradients of 
the host galaxies as reported in Table 3, column (7). With the exception of 
PKS 2254+074 (see Appendix), all colour profiles show a modest colour gradient 
(average $\Delta$(R--H)/$\Delta$(log r) = --0.09$\pm$0.04) that makes the 
galaxies bluer in the outer regions. The sign and amplitude of the color 
gradient is similar to that exhibited by normal non-active ellipticals 
($\Delta$(V--K)/$\Delta$(log r) = --0.16$\pm$0.18; Peletier, Valentijn \& 
Jameson 1990; 12 ellipticals); $\Delta$(V--K)/$\Delta$(log r) = 
--0.26$\pm$0.15; Schombert \etal 1993; 16); the R--H gradient tends to be 
smaller than that of V--K, probably due to the smaller wavelength baseline. 
Finally, we show in Fig. 4 the colour--magnitude diagram of the BL Lac host 
galaxies, compared with those of elliptical and S0 galaxies in the Virgo and 
Coma clusters (Bower, Lucey \& Ellis 1992a,b). With the exceptions of PKS 
0548--322 and PKS 2254+074 (see Appendix for details), the BL Lac hosts 
follow reasonably well the established relation for elliptical galaxies. 

\begin{figure}
\psfig{file=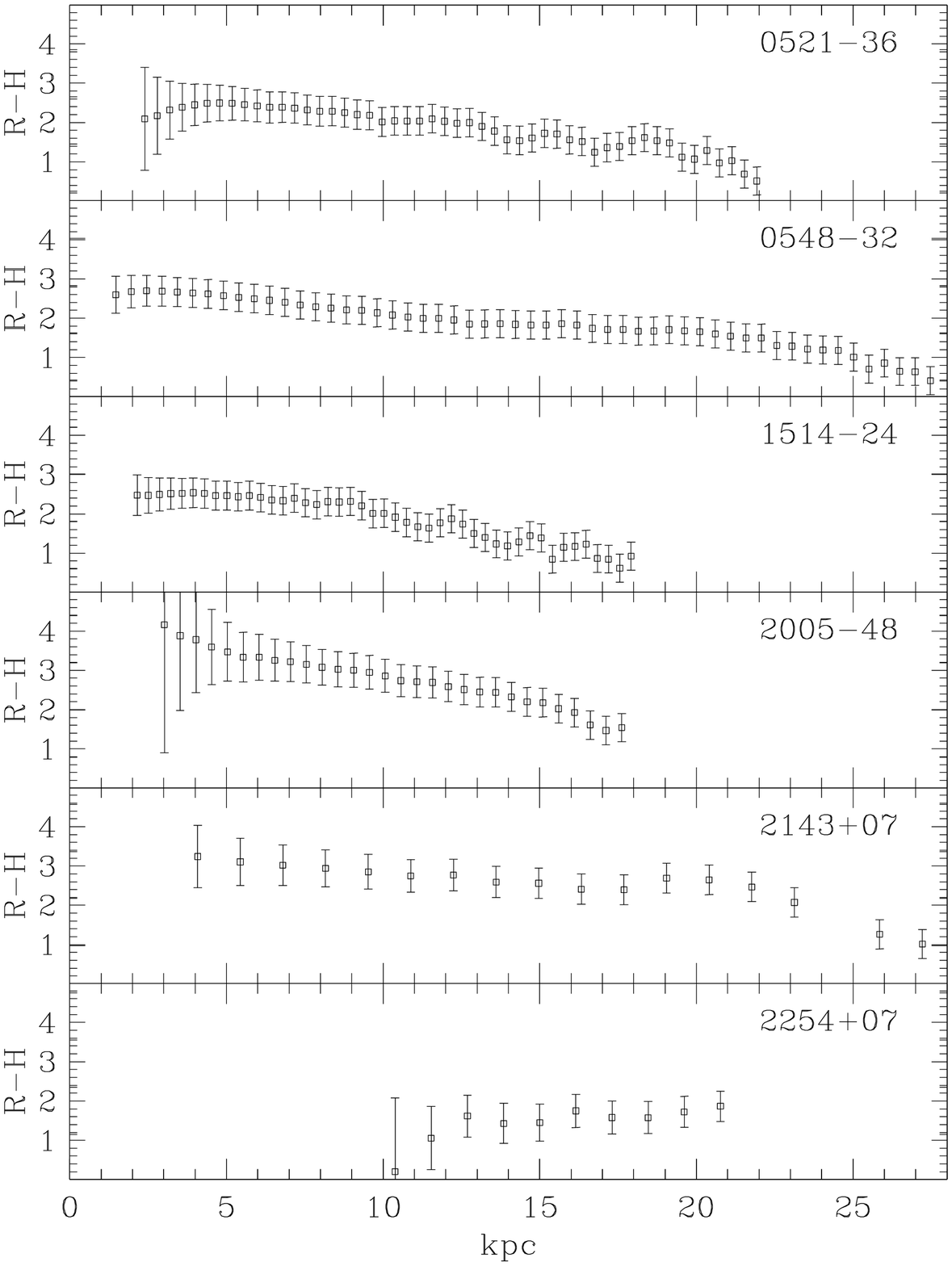,width=9cm,height=13cm}
\caption{\label{fig:fig3} 
The host galaxy R--H colour vs. radius for the resolved BL Lacs in the sample.}
\end{figure}

\begin{figure}
\psfig{file=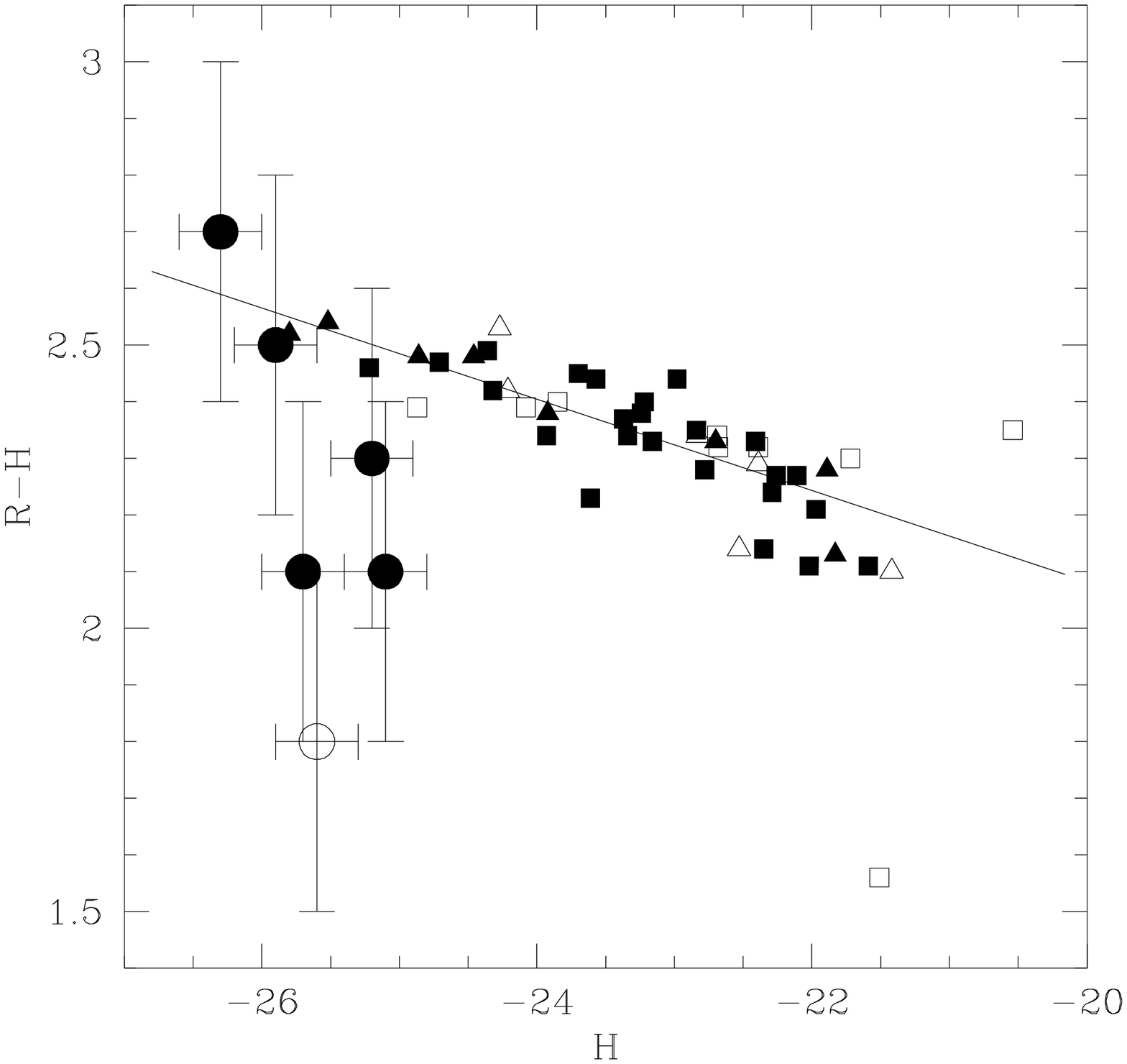,width=9cm,height=9cm}
\caption{\label{fig:fig4} 
The R--H vs. H colour--magnitude diagram for the BL Lac host galaxies (filled 
circles: Table 3, column 6) in this sample. The uncertain value for PKS 
2254+074 (see Appendix) is indicated as an open circle. The other symbols 
indicate elliptical galaxies in the Virgo (filled triangles) and Coma (filled 
squares) clusters, and S0 galaxies in the Virgo (open triangles) and Coma 
(open squares) clusters (from Bower \etal 1992a,b). The solid line shows the 
best--fit regression line for Virgo and Coma clusters (Bower \etal 1992b). 
The V--K vs. V diagram of Bower \etal (1992b) has been transformed into R--H 
vs. H assuming V--R = 0.6, H--K = 0.2 (see text), and distance moduli for 
Virgo and Coma clusters m--M = 31.0 and 34.6, respectively (Bower \etal 
1992b).}
\end{figure}

There is increasing evidence that the photometric properties of elliptical 
galaxies (at least of those in nearby clusters) can be explained in terms of 
a single burst of star formation at high redshift, followed by passive 
stellar evolution (\eg Stanford, Eisenhardt \& Dickinson 1998 and references 
therein). The similarity of the colours and colour gradients of BL Lac hosts 
to those of normal elliptical galaxies then indicates that the nuclear 
activity has only marginal effect, if any, on the overall properties of the 
host galaxy. For example, galaxy interactions and mergers, often invoked to 
explain the BL Lac phenomenon and AGN in general (\eg Heckman 1990), do not 
seem to induce strong recent star formation in the BL Lac hosts. This is 
perhaps due to a very low gas density in the host and/or in the interacting 
galaxy. Alternatively, it may be that the timescales of the star formation 
episodes and nuclear activity are different. 

Table 5 presents a comparison of the H--band absolute magnitudes of the BL 
Lac hosts with the average values of relevant samples of blazars and RGs from 
previous optical and NIR studies in the literature. Column (1) gives the 
sample; (2) the filter; (3) the number of objects in the sample; (4) the 
average redshift of the sample; and (5) and (6) the average H band nuclear 
and k-corrected host galaxy absolute magnitude of the sample. All magnitudes 
were transformed into our adopted cosmology and into the H band. Based on the 
similarity of the BL Lac host galaxies to giant ellipticals (see discussion 
above), the transformation of published magnitudes to the H band was obtained 
assuming the following colours of giant ellipticals, computed from the 
literature values given in Table 4: V--H = 3.1, R--H = 2.5, I--H = 1.8, and 
H--K = 0.2. Finally, transformation from Gunn r filter (r--H = 2.8) was 
obtained from R--H = 2.5 (above) and r--R = 0.3, appropriate for low redshift 
ellipticals (Fukugita, Shimasaku \& Ichikawa 1995).  

\begin{table*}
\begin{center}
\begin{tabular}{llrlll}
\multicolumn{6}{c}{{\bf Table 5.} Comparison of average host galaxy 
properties with other samples.}\\
\hline\\
\multicolumn{1}{c}{Sample$^a)$} & filter & N & $<z>$ & $<M_H(nuc)>$ & 
$<M_H(host)>^b)$ \\
\hline\\
(1) & (2) & (3) & (4) & (5) & (6) \\
\hline\\
BL this work & H & 7 & 0.112$\pm$0.068 & -25.7$\pm$1.7 & -25.8$\pm$0.5 \\
	     &	 &   &                 &               &             \\
L* Mobasher \etal (1993) & K & 136 & 0.077$\pm$0.030 & & -25.0$\pm$0.2 \\
	     &	 &   &                 &               &             \\
BCM Thuan \& Puschell (1989) & H & 84 & 0.074$\pm$0.026 & & -26.3$\pm$0.3 \\
BCM Aragon-Salamanca \etal (1998) & K & 25 & 0.449$\pm$0.266 & & 
-27.0$\pm$0.3\\
	&	&   &       &         &             \\
BL Falomo (1996) & R & 11 & 0.143$\pm$0.082 & -26.3$\pm$2.1 & -26.0$\pm$0.6\\
BL Wurtz \etal (1996) & r & 35 & 0.266$\pm$0.162 & -24.5$\pm$2.3 & 
-26.3$\pm$0.7\\
BL Falomo \etal (1997b), Jannuzi \etal (1997) & I & 6 & 0.351$\pm$0.172 & 
-26.0$\pm$0.8 & -26.0$\pm$0.8 \\
	&	&   &       &         &             \\
RG Fasano \etal (1996)         & R & 25 & 0.061$\pm$0.027 & & -26.6$\pm$0.6\\
RG low--z Zirbel (1996)  & V & 55 & 0.107$\pm$0.047 & &-25.8$\pm$0.6\\
RG F--R II high--z Zirbel (1996) & V & 25 & 0.389$\pm$0.053 & & -25.9$\pm$0.4\\
RG F--R II Taylor \etal (1996) & K & 12 & 0.214$\pm$0.049 & -25.1$\pm$0.7 & 
-26.1$\pm$0.8 \\
	&	&   &       &         &             \\
FSRQ/R+M KFS98 & H & 9 & 0.671$\pm$0.157 & -29.7$\pm$0.8 & -26.7$\pm$1.2 \\
FSRQ/R KFS98 & H & 4 & 0.673$\pm$0.141 & -30.2$\pm$0.7 & -27.8$\pm$0.3 \\
\hline\\
\multicolumn{6}{l}{$^a)$: BL = BL Lac objects; L$^*$ = typical field galaxies;
BCM = brightest cluster members; RG = radio}\\ 
\multicolumn{6}{l}{galaxies; FSRQ = flat spectrum radio quasars.}\\ 
\multicolumn{6}{l}{$^b)$: Transformation to H band was done assuming the 
following galaxy colours: V--H = 3.1, r--H = 2.8,}\\
\multicolumn{6}{l}{R--H = 2.5, I--H = 1.8, and H--K = 0.2 (for references, 
see text and Table 4).}\\
\end{tabular}\\
\end{center}
\end{table*}

In Fig. 5 (upper panel) we show the H--z (Hubble) diagram for the BL Lac 
hosts (this work), together with data for FSRQs (KFS98) and F--R II RGs 
(Taylor \etal 1996), compared to the established relation for RGs (solid 
line; \eg Lilly \& Longair 1984; Lilly, Longair \& Miller 1985; Eales \etal 
1997), and the evolutionary model for elliptical galaxies derived from 
passive stellar evolution models (Bressan, Chiosi \& Fagotto 1994, dashed 
line), normalized to the average redshift and magnitude of the low redshift 
RGs from Taylor \etal 1996). The resolved BL Lac hosts lie quite well on the 
H--z relation. In Fig. 5 (lower panel) we show the H--z diagram for the mean 
values of various blazar and RG samples from the literature. These values 
consistently follow the H--z relationship (within the errors), with the 
average value for the BL Lacs derived in this work well matching the 
relationship. 

\begin{figure}
\psfig{file=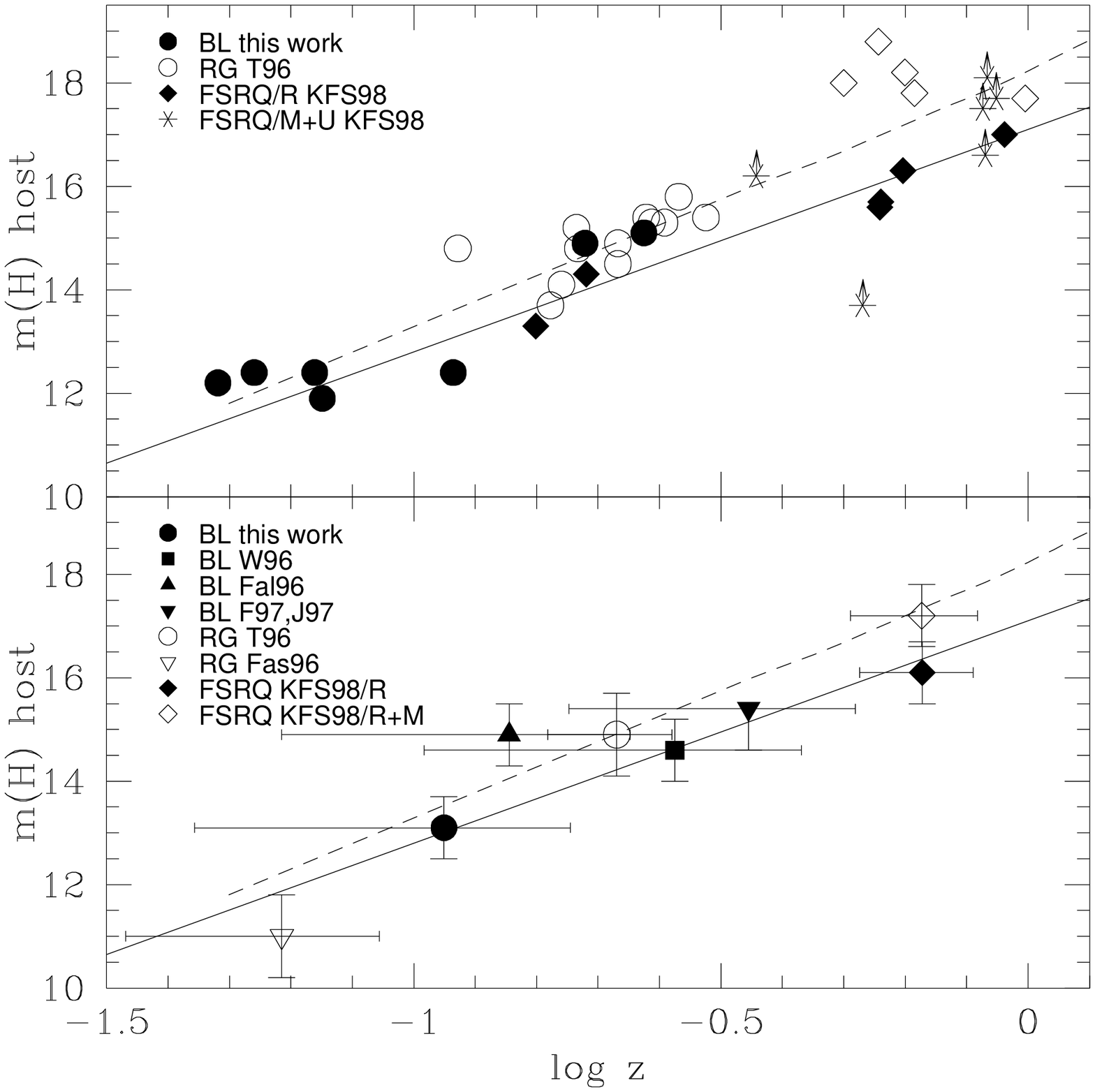,width=9cm,height=9.8cm}
\caption{\label{fig:fig5} 
{\bf a)} Plot of the apparent H magnitude of the host galaxies vs. redshift 
(``Hubble diagram''). The BL Lacs are marked as filled circles, the FSRQs 
from KFS98 (H band) as filled diamonds (resolved hosts) and asterisks 
(marginally resolved and unresolved hosts), and RGs from Taylor \etal (1996; 
K band) as open circles. The solid line is the K-z relation for RGs derived 
by Lilly \etal (1985) and Eales \etal 1997), converted to H magnitudes 
assuming H--K = 0.2, and the dashed line is the evolutionary model for 
elliptical galaxies derived from passive stellar evolution models (Bressan 
\etal 1994), normalized to the average redshift and magnitude of the low 
redshift RGs from Taylor \etal (1996). 
{\bf b)} As Fig. 5a, except we show the mean values of our sample in 
comparison with samples from literature. Symbols are as in Fig. 5a, except 
the open diamonds are a combined sample of the resolved and marginally 
resolved FSRQs. Additional samples based on optical imaging of BL Lacs from 
Wurtz \etal (1996), Falomo (1996), Falomo \etal (1997b) and Jannuzi \etal 
(1997) and of RGs from Fasano, Falomo \& Scarpa (1996) are marked as 
indicated in the figure.}
\end{figure}

In Fig. 6 (upper panel), we show the H--band host galaxy absolute magnitude 
vs. redshift for the BL Lacs (this work), FSRQs (KFS98) and RGs (Taylor \etal 
1996). In Fig. 6 (lower panel), the same diagram is shown for the mean values 
of various blazar and RG samples. The average H--band absolute magnitude of 
the seven clearly resolved low redshift (z$\leq$0.2) BL Lac hosts is M(H) = 
--25.8$\pm$0.5, and the average and median bulge scalelength R(e) = 
8.8$\pm$9.9 kpc and 5.0 kpc. The host galaxies are therefore large (all have 
R(e)$\geq$3 kpc, the upper boundary of normal local ellipticals; Capaccioli, 
Caon \& D'Onofrio 1992) and slightly more luminous than an L$^*$ galaxy (M(H) 
= --25.0$\pm$0.2; Mobasher \etal 1993). Indeed, we find no BL Lac host fainter 
than L$^*$. As found in the optical by Wurtz \etal (1996), the BL Lac hosts 
have slightly, but not significantly lower luminosities in the NIR than 
nearby brightest cluster member galaxies (BCM; z = 0.07$\pm$0.03; M(H) = 
--26.3$\pm$0.3; Thuan \& Puschell 1989). However, they appear  significantly 
fainter than BCMs at higher redshift (z = 0.45$\pm$0.27; M(H) = 
-27.0$\pm$0.3; Aragon-Salamanca, Baugh \& Kauffmann 1998). 

\begin{figure}
\psfig{file=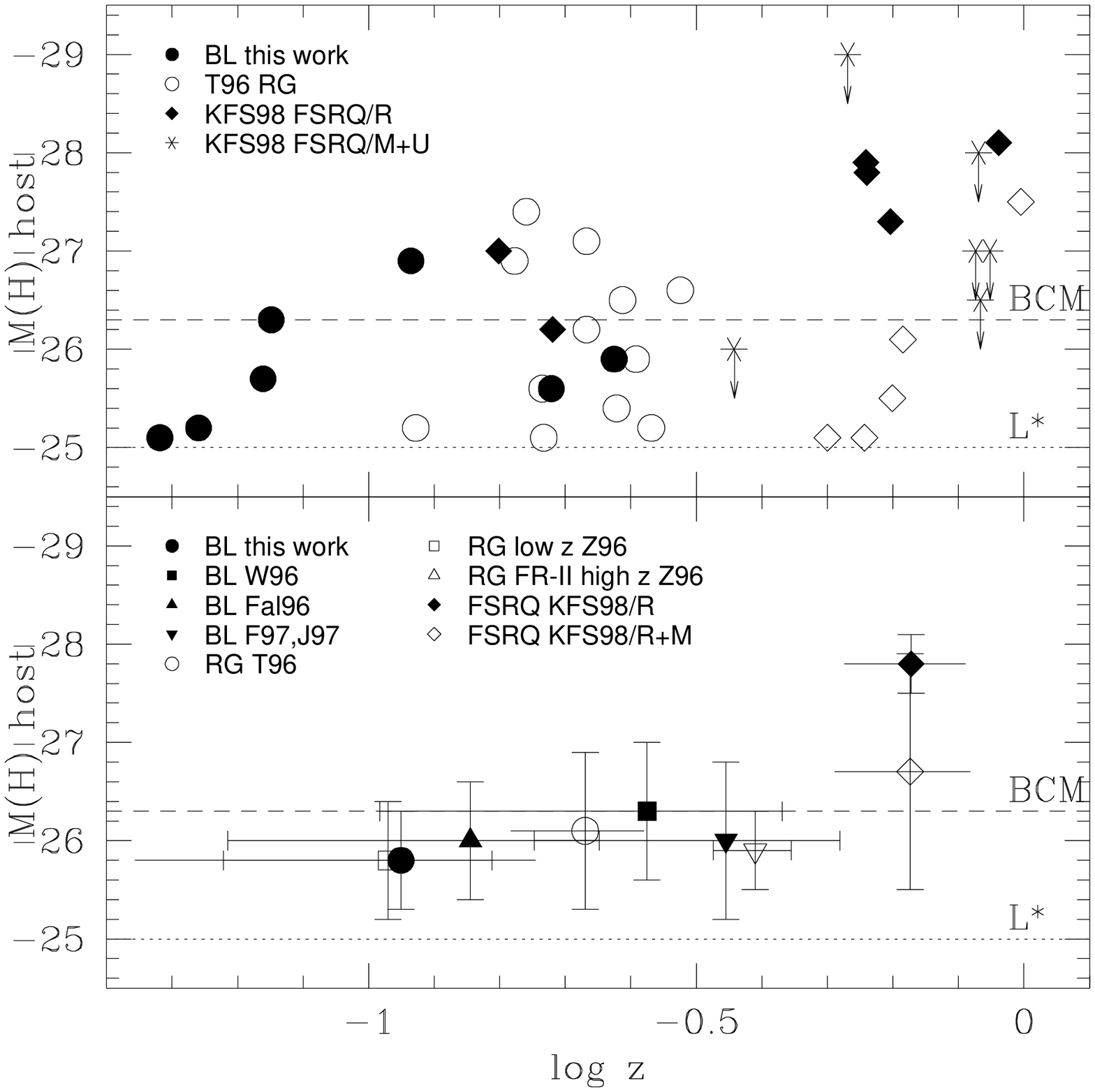,width=9cm,height=9.7cm}
\caption{\label{fig:fig6} 
{\bf a)} Plot of the absolute H band magnitude of the host galaxies vs. 
redshift. For symbols, see Fig. 5a. The dotted lines are the luminosities of 
L* (M(H)$\sim$-25.0; Mobasher, Sharples \& Ellis 1993) and brightest cluster 
member galaxies (BCM; M(H)$\sim$-26.3; Thuan \& Puschell 1989).
{\bf b)} As Fig. 6a, except for the mean values of various samples. For 
symbols, see Fig. 5 and 6a. Additional samples of RGs from Zirbel (1996) are 
marked as indicated in the figure.}
\end{figure}

Although the number of objects in this study is small, it is interesting to 
compare the NIR absolute magnitudes of BL Lac hosts to those of available 
samples of BL Lacs, RGs and FSRQs from the literature. These samples span a 
moderately large range in redshift from z$\sim$0.03 up to z$\sim$0.7. The 
average host galaxy magnitudes for the various samples are given in Table 5, 
and shown in Fig. 6. The BL Lacs in our study have host galaxies very similar 
in luminosity to those found by previous optical studies of BL Lacs if normal 
colors are assumed. Interestingly, there is good agreement between the BL Lac 
hosts in this study and the F-R II RGs of Taylor \etal (1996), the only RG 
sample studied in the NIR and thus almost free from colour term uncertainty. 
The lack of direct comparison sample of F-R I RGs, however, prevents us from 
discussing the issue of the parent population of BL Lac objects further.  

Since BL Lacs share many properties (\eg variability and polarization) with 
FSRQs, it is interesting to compare the host properties of these two types of 
blazars. Recently, KFS98 found for clearly resolved FSRQ hosts at z$\sim$0.7 
M(H) = --27.8$\pm$0.3, while the value adding their marginally resolved hosts 
is M(H) = --26.7$\pm$1.2 (Table 5). Because not all FSRQs were resolved, it 
is possible that the average host luminosity of FSRQs is even fainter. 
Moreover, given the large difference of redshift between the BL Lac and FSRQ 
samples, it is also possible that the $\sim$0.5 mag difference in host 
luminosity is due to cosmological evolution in the stellar population of an 
elliptical galaxy that makes it fade by $\sim$0.8 mag between z$\sim$1 and 
z$\sim$0 (\eg models by Bressan \etal 1994). Indeed, there is increasing 
observational evidence for significant evolution in the colours and 
luminosity of early--type galaxies (\eg Aragon-Salamanca \etal 1998 and 
references therein). The present data suggests that BL Lac hosts are less 
luminous than those of FSRQs but, given the limits described above, 
additional homogeneous observations of BL Lacs and FSRQs are needed to reach 
a firm conclusion.

\subsection{The nuclear component}

The average absolute magnitude of the fitted nuclear component for the 
clearly resolved BL Lacs is M(H) = --25.7$\pm$1.7. On the other hand, the 
FSRQs have much brighter nuclear component (M(H) = --29.7$\pm$0.8; KFS98). 
This difference in the strength of the nuclear component is also more evident 
considering the nucleus/galaxy luminosity ratio L(nuc)/L(gal) in Table 2. 
None of the low redshift BL Lacs (or the RGs from Taylor \etal 1996) have 
L(nuc)/L(gal)$\geq$10, whereas about half of the FSRQs are above this limit 
(KFS98). From Figs. 5 and 6 it is clear that the host galaxies of the various 
samples considered  here are not dramatically different in intrinsic 
luminosity, especially if cosmological evolution is taken into account. 
Therefore, this difference suggests that FSRQs exhibit a nuclear component 
which is systematically brighter than that of the low redshift BL Lac 
objects, possibly due to either intrinsically higher luminosity or a larger 
Doppler beaming factor. While this result disagrees with the current unified
models of AGN, further observations of larger and unbiased samples of BL Lacs 
and FSRQs are needed to elucidate this point.

\section{Conclusions}

In this paper we have presented the results of a near--infrared imaging study 
of a sample of 11 BL Lac objects, for most of which the host galaxy is 
clearly resolved. Consistently with what is found in optical studies, we find 
that the host galaxies of low redshift BL Lacs are large (average bulge scale 
length R(e) = 8.8$\pm$9.9 kpc) and luminous (average M(H) = --25.8$\pm$0.5); 
they are more luminous than L$^*$ galaxies (by $\sim$1 mag) but of similar 
luminosity to or slightly fainter than the brightest cluster galaxies. Our 
NIR study was able for the first time to address the issue of the 
optical--NIR colour of BL Lac host galaxies. The average R--H colour and 
colour gradient of the BL Lac hosts are consistent with those of non-active 
early-type galaxies, suggesting that the nuclear activity does not have much 
effect on the star formation history of the host galaxies. The 
nucleus--to--galaxy ratio of BL Lacs is similar to that found in low redshift 
RGs and consistent with what found in previous optical studies of BL Lacs. 
However, it is smaller that that found for the higher redshift FSRQs (KFS98), 
suggesting there is a difference in the intrinsic brightness of the nuclear 
source or in the Doppler beaming factor between the two types of blazars. We 
finally encourage a systematic NIR multiwavelength study of a large, well 
defined sample of BL Lac objects and their immediate environments with the 
new generation large NIR arrays. 

\section*{Acknowledgments} 
JKK acknowledges a research grant from the Academy of Finland during the 
initial part of this work. We thank the anonymous referee for prompt 
comments. This research has made use of the NASA/IPAC Extragalactic Database 
(NED), which is operated by the Jet Propulsion Laboratory, California 
Institute of Technology, under contract with the National Aeronautics and 
Space Administration.

\section*{Appendix: Notes on individual objects and comparison with previous 
NIR photometry}

{\bf PKS 0048--097.} Our H--band photometry agrees well with previous studies 
(Table 6). High resolution optical imaging by Falomo (1996) revealed a faint 
(m(R)$\sim$22.5) companion object $\sim$2.5$''$ to the E of the BL Lac 
nucleus. The host galaxy of PKS 0048--097 remained unresolved by Falomo 
(1996), but assuming an elliptical host of M(R) = --23.5, he derived a lower 
limit to the redshift of z$\geq$0.5. The host galaxy remains unresolved in 
the NIR. Some hint of the eastern companion object is visible in NIR.

\begin{table}
\begin{center}
\begin{tabular}{llll}
\multicolumn{4}{c}{{\bf Table 6.} H--band photometry compared to}\\ 
\multicolumn{4}{c}{previous literature photometry.}\\
\hline\\
\multicolumn{1}{c}{Name} & H mag$^a)$ & H mag range & References \\
\hline\\
\multicolumn{1}{c}{} & (this work) & (literature) & \\
\hline\\
PKS 0048--097 & 13.64 & 12.78 - 14.86 & M90, A82\\
PKS 0118--272 & 13.29 & 12.69 - 13.10 & M90, B92\\
PKS 0521--365 & 11.75 & 10.80 - 12.64 & A74, B86\\
PKS 0537--441 & 13.04 & 11.69 - 13.34 & T86, B92\\
PKS 0548--322 & 13.39 & 12.81 - 13.13 & G79, A82\\
PKS 1514--241 & 12.47 & 11.54 - 12.58 & B92, B86\\
PKS 1538$+$14 & 14.57 & 14.23 - 14.67 & B86, G93\\
PKS 2005--48  & 11.34 & 11.40	  & B92\\
MS 2143.4$+$07 & 15.25 & 14.60	  & G93\\
PKS 2155--305 & 10.93 & 10.70 - 11.06 & B86, B92\\
PKS 2254$+$074  & 14.38 & 12.99 - 14.09 & A82, M90\\
\hline\\
\multicolumn{4}{l}{$^a)$ Aperture photometry in a 6 $''$ diameter aperture.}\\ 
\multicolumn{4}{l}{A74 = Andrews, Glass \& Hawarden (1974); A82 = Allen,}\\
\multicolumn{4}{l}{Ward \& Hyland (1982); B86 = Brindle \etal (1986); B92 =}\\
\multicolumn{4}{l}{Bersanelli \etal (1992); G79 = Glass (1979); G93 = Gear}\\
\multicolumn{4}{l}{(1993); M90 = Mead \etal (1990); T86 = Tanzi \etal (1986)}\\
\end{tabular}\\
\end{center}
\end{table}

{\bf PKS 0118--272.} Our H--band magnitude is slightly fainter than that 
found in the literature (Table 6). The host galaxy remains unresolved in the 
optical (Falomo 1996), however, this is consistent with the presence of a 
luminous host (M(R) = --23.5) at the proposed redshift of z$\geq$0.559. The 
host galaxy remains unresolved in the NIR. 

{\bf PKS 0521--365} is at redshift z = 0.055. Our H-band photometry agrees 
well with literature values (Table 6). The host galaxy has been imaged in the 
optical by Falomo (1994), who found best fit for a giant elliptical (M(R) = 
--23.2 and R(e) = 9 kpc) with a faint stellar disk, and by Wurtz \etal 
(1996), who fit the host with an elliptical galaxy having M(R) = --23.0 and 
R(e) = 5.4 kpc. These values are in good agreement with those found in this 
study (M(H) = --25.4, R(e) = 5.8 kpc). 

{\bf PKS 0537--441} is a high redshift BL Lac (z = 0.896; Peterson \etal 
1976). Our H--band magnitude agrees well with previous studies (Table 6). The 
existence and nature of extended emission around this BL Lac, and its 
relevance to gravitational lensing, has been debated in the literature (\eg 
Stickel, Fried \& K\"{u}hr 1988; Falomo, Melnick \& Tanzi 1992). No 
nebulosity surrounding this high redshift BL Lac is detected in the NIR, in 
agreement with Falomo \etal (1992), who conclude that there is no evidence 
for a foreground lensing galaxy. 

{\bf PKS 0548--32} is the dominant member of a rich cluster of galaxies at 
redshift z = 0.069 (Fosbury \& Disney 1976). Our H--band photometry is 
slightly fainter than found in previous studies (Table 6). Falomo \etal 
(1995) fit the host galaxy with a giant elliptical (M(R) = --24.2 and R(e) = 
51 kpc) and a faint stellar disk. Wurtz \etal (1996) found for the host 
galaxy an elliptical fit, with M(R) = --23.2 and R(e) = 14 kpc. The values 
derived by us in the NIR are M(H) = --25.7, and R(e) = 8.3 kpc. Note that this
BL Lac is surrounded by an extended halo, which was detected in the optical 
by Falomo \etal (1995), but not in the NIR. Therefore, the real colour of the 
host is redder than that derived from difference of total magnitudes in the 
two bands (see table 3).

{\bf PKS 1514--241 = AP Lib} is at redshift z = 0.0486 (Disney, Peterson \& 
Rodgers 1974). Our H--band magnitude agrees reasonably well with literature 
photometry (Table 6). The elliptical host galaxy has been studied extensively 
in the optical. Baxter \etal (1987) obtained M(V) = --22.8, Abraham \etal 
(1991) M(R) = --22.8 and R(e) = 7.5 kpc, and Stickel \etal (1993) derived 
M(R) = --23.5 and R(e) = 11.5 kpc. Finally, we have analysed our unpublished 
optical images of this BL Lac, obtained at the ESO 2.2m telescope, and find 
M(R) = --22.9 and R(e) = 4.3 kpc. The host galaxy parameters derived by us in 
the NIR (M(H) = --25.1, R(e) = 3.3 kpc) are in reasonable agreement with 
those found in the previous studies. 

{\bf PKS 1538+149} is at redshift z = 0.605 (Stickel \etal 1993). Our H-band 
photometry agrees well with those found in the literature (Table 6). Stickel 
\etal (1993) could not resolve the host galaxy, while Wurtz \etal (1996) 
found a marginal fit for the host, with M(R) = --24.2 and R(e) = 12 kpc. 
Falomo \etal (1997b) derived for the host from HST imaging M(I) = --25.2. The 
host galaxy remains unresolved in the NIR. 

{\bf PKS 2005--489} is at redshift z = 0.071 (Falomo \etal 1987). Our H--band 
magnitude agrees well with previous studies (Table 6). Stickel \etal (1993) 
derived for the host galaxy M(R) = --24.2 and R(e) = 5.2 kpc, while Falomo 
(1996) found the host to be an elliptical with M(R) = --23.7 and R(e) = 11 
kpc. The absolute magnitude derived by us in the NIR is: M(H) = --26.3.

{\bf MS 2143.4+07} is at redshift z = 0.237. Our H--band photometry is 
slightly fainter than found in previous studies (Table 6). Wurtz \etal (1996) 
derived for the elliptical host galaxy M(R) = --23.4 and R(e) = 12 kpc, while 
Jannuzi \etal (1997) derived from HST imaging M(I) = --24.0 and R(e) = 9.0 
kpc. These values are in excellent agreement with those derived by us in the 
NIR (M(H) = --25.9, R(e) = 5.4 kpc). 

{\bf PKS 2155-305} is at redshift z = 0.116 (Falomo, Pesce \& Treves 1993) 
and it is one of the brightest and most studied BL Lacs and is often 
considered the prototype of X--ray selected BL Lacs. Our H--band magnitude is 
in good agreement with literature values (Table 6). For the host galaxy, 
Falomo \etal (1991) derived M(R) = --24.4 and R(e) = 13 kpc, while Wurtz 
\etal (1996) could not resolve the host, with M(R)$\geq$--23.1. The NIR 
properties of the host (M(H) = --26.8, R(e) = 5.7 kpc) are in good agreement 
with Falomo \etal (1991).  The companion galaxy at 4.2$''$ E of PKS 2155-305, 
previously detected in the optical (Falomo \etal 1991), is clearly seen in 
the NIR image (see Fig. 1).

{\bf PKS 2254+074} is at redshift z = 0.190 (Stickel \etal 1993). We obtained 
only a short exposure of this source under poor sky conditions. We also note 
that the BL Lac was situated in a bad area of the array during the 
observations. Consequently, both the NIR photometry (slightly fainter than 
found in literature; Table 6), and the host properties are less accurate than 
for the other resolved BL Lacs. Optical determinations of the host galaxy 
have yielded M(R) = --24.1 and R(e) = 14.5 kpc (Stickel \etal 1993), M(R) = 
--23.9 and R(e) = 15 kpc (Falomo 1996), M(R) = --23.9 and R(e) = 17 kpc 
(Wurtz \etal 1996) and M(I) = --24.8 and R(e) = 15 kpc (Falomo \etal 1997b). 
The absolute magnitude derived by us in the NIR (M(H) = --25.5) is $\sim$1 
mag fainter than implied by the optical studies, assuming normal elliptical 
colours. 

\end{document}